\documentclass[aps,prb,twocolumn,superscriptaddress]{revtex4-1}
\usepackage{graphicx}
\usepackage{picture}
\usepackage{xcolor}
\usepackage{color}
\usepackage{tikz}
\usepackage{amsmath}
\usepackage{mathrsfs}

\begin{document}

\title{Pulse-Enhanced Two-Photon Interference with Solid State Quantum Emitters
}

\author{Herbert F Fotso }
\affiliation{Department of Physics, University at Albany (SUNY),  Albany, New York 12222, USA}

\begin{abstract}
The ability to entangle distant quantum nodes is essential for the construction of quantum networks and for quantum information processing. For solid-state quantum emitters used as qubits, it can be achieved by photon interference. When the emitter is subject to spectral diffusion, this process can become highly inefficient, impeding the achievement of scalable quantum technologies. We study two-photon interference in the context of a Hong-Ou-Mandel (HOM)-type experiment for two separate quantum emitters, with different detunings with respect to a specific target frequency. We evaluate the second order coherences that characterize photon indistinguishability between the two emitters. We find that the two-photon interference operation that is inefficient in the absence of a control protocol, when the two detunings are different and spectral overlap is lessened, can be highly improved by a periodic sequence of $\pi$ pulses at a set target frequency. Photon indistinguishability in solid state and other quantum emitters subject to spectral diffusion can thus be enhanced by the proposed  pulse sequence and similar external control protocols despite the fluctuations in the environment.
\end{abstract}

\maketitle

\section{Introduction}
Recent progress in the field of Quantum Information Processsing (QIP) has used a variety of platforms as qubits.
Several QIP experimental milestones have been reached using as quantum bits quantum-dots\cite{Imamoglu_Awschalom_QDOT_PRL_99, SantoriVuckovicYamamoto_QDOT_Nat2002, Kuhlmann_Warburton_QDOT_NatPhys2013, Gao_Imamoglu_NatComm2013}, nitrogen-vacancy (NV) centers\cite{Humphreys_Hanson_NV_entanglement, ChildressRepeater, Bernien_Hanson_Nature2013, Hanson_loopholeFree_Nature2015} or silicon-vacancy(SiV) centers\cite{Sipahigil_HOM_SiV_2014, Rogers_SiV} in diamond, trapped ions\cite{Maunz_Monroe_HOMIons} or neutral atoms\cite{Rosenfeld_etal_BellTest_atoms, Beugnon_Grangier_HOMAtoms} and superconducting circuits\cite{Chou_Schoelkopf_2018, Kurpiers_Wallraff_2018, Axline_Schoelkopf_2018, CIbarcq_Devoret_2018, Zhong_Cleland_2019}. In particular, experiments have achieved teleportation of quantum states and loophole free Bell inequality tests using quantum bits separated by macroscopic distances\cite{Hanson_loopholeFree_Nature2015, Humphreys_Hanson_NV_entanglement, Gao_Imamoglu_NatComm2013, Pfaff_Hanson_Science2014}. A central component of these experiments as well as the implementation of two-qubit gates is the two-photon interference that is essential for the generation of entanglement between distant quantum bits. This is typically achieved by interfering two photons emitted by the qubits to be entangled on a beam splitter in a Hong-Ou-Mandel (HOM)-type experiment\cite{HOM}. The success of this operation is in turn tied to the indistinguishability of the photons. For solid state quantum emitters and other systems in dynamic environments, the emission/absorption spectrum can drift in an uncontrolled way away from a target frequency as a result of fluctuations in the surrounding bath (e.g. changes in local strain and motion of neighboring charges). These variations in spatial, temporal and spectral profiles can compromise photon indistinguishability and significantly hamper entanglement generation between distant quantum bits and other photon-mediated processes essential to scalable quantum networks and QIP\cite{AwschalomHansonZhou_opticsQIP, AtatureEnglund_Wrachtrup_NatRevMat_2018, HansonAwschalom_QIP_ss_08, NV_Review_PhysRep2013, CarterGammonQDcavity, JeffKimble_qtmInternet, NielsenChuangBook}.

This problem continues to receive a great deal of attention both from the point of view of material design and from the point of view of external control\cite{AmbroseMoerner_spectralDiffusion_Nature1991, Schroeder_Englund_SPE_NatComm2017, QDOTs_HOM_Atature, Aharonovich_Englund_Toth_SS_SPE, Fu_Beausoleil_PRL2009, Acosta_Beausoleil_PRL2012, Dreau_Jacques, Pfaff_Hanson_Science2014, Basset_Awschalom_PRL2011, FaraonEtalNatPhot, Hansom_Atature_APL2014, Crooker_Bayer_PRL2010, Calajo_Passante_PRA2017, JSLee_Khitrin_JPhysB2008, IDS_2017}. 
In one approach proposed in recent work,  it was shown that appropriate pulse sequences can be applied to quantum emitters in a dynamic environment to produce an emission or absorption spectrum that has little dependence on the fluctuations in the environment\cite{FotsoEtal_PRL2016, FotsoDobrovitski_Absorption, Fotso_JPhysB2019}. It was, for instance, shown that a  periodic sequence of $\pi$-pulses can maintain the bulk of the emission spectrum at a central peak located at the pulse carrier frequency and satellite peaks at frequencies shifted from this central peak by integer multiples of $\pm \pi/\tau$; where $\tau$ is the period of the pulse sequence. This lineshape is unchanged for various detunings as long as the pulse sequences are appropriately adjusted\cite{FotsoEtal_PRL2016,Fotso_JPhysB2019}. In the context of quantum information processing, this naturally raises the question of how different quantum emitters, each with their own dynamic environment, would fare in a HOM-type two-photon interference when they are driven by such a pulse sequence. 

The goal of this paper is to answer this question. Namely, we consider the problem of photon indistinguishability for two distant quantum emitters in dynamic environments. We evaluate the intensity correlation at the detectors in a HOM-type two-photon interference when the emitters have different detunings $\Delta_1$  and $\Delta_2$ with respect to the pulse carrier frequency of an applied periodic sequence of $\pi$-pulses with period $\tau$. We find that in the situation without control protocol, when  $\Delta_1$ and $\Delta_2$ are significantly different, the intensity correlation exhibits beating with vanishing values periodically as a function of the delay time $\theta$ between the two the detectors. On the contrary, in the presence of the pulse sequence, the intensity correlation vanishes at time delay $\theta = 0$ but keeps a finite value for finite $\theta$. This corresponds to enhanced photon indistinguishability between two qubits that originally have different environments and spectral profiles.

\section{Two-Photon Interference, Spectral Diffusion: Model}
\begin{figure}[htbp]
\includegraphics[height=7.0cm, width=8.0cm]{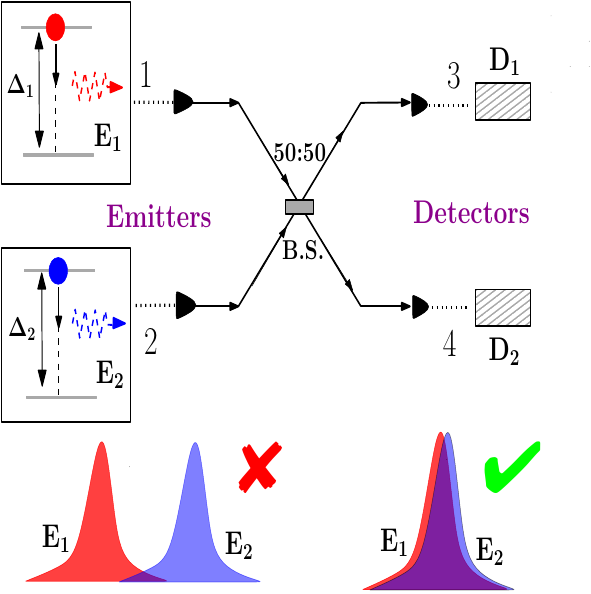}
\caption{Two Photon Interference from two distant quantum emitters. Photons from Emitters $E_1$ and $E_2$ at spacetime locations $1$ and $2$( respective detunings $\Delta_1$ and $\Delta_2$ measured in the frame rotating at the set target frequency $\omega_0$) are sent to input ports of a 50:50 beam splitter and then measured at detectors $D_1$ and $D_2$ at spacetime locations $3$ and $4$. Indistinguishable photons will coalesce and emerge at the same port (top). This indistinguishability is synonymous with similar spatial  temporal and spectral profiles. For high efficiency, the entanglement operation requires overlapping spectra whereas it is inefficient for low spectral overlap (bottom).}
\label{fig:HOM}
\end{figure}

The two-photon interference, pictured in Fig.\ref{fig:HOM}, involves two separate and independent quantum emitters. Photons from Emitters $E_1$ and $E_2$, with respective detunings $\Delta_1$ and $\Delta_2$ measured in the frame rotating at the set target frequency $\omega_0$, at spacetime locations $1$ and $2$, are sent to input ports of a 50:50 beam splitter and then measured at detectors $D_1$ and $D_2$, located at spacetime locations $3$ and $4$ beyond the output ports of the beams splitter. Indistinguishable photons will coalesce and emerge at the same port (Fig.\ref{fig:HOM} top). This indistinguishability is synonymous with identical spatial, temporal and spectral profiles. Thus, for high efficiency, the entanglement operation requires overlapping spectra whereas it is inefficient for low spectral overlap (Fig.\ref{fig:HOM} bottom).

We model each individual solid-state emitters as a two-level systems coupled to a radiation bath. The two-level system emits a photon in the course of a spontaneous transition from its excited state $|e\rangle$, located at the energy $\hbar\omega_1$ above the ground state $|g\rangle$ (below we  will set $\hbar=1$). The optical control pulses, each of very short duration $t_p$, are applied at the pulse carrier frequency, $\omega_0$, at appropriate times. It is thus convenient to work in the rotating-wave approximation (RWA), using the basis rotating at frequency $\omega_0$. The system corresponding to the driven emitter in the radiation bath can then be described by a Hamiltonian of the form:


\begin{eqnarray}
H = \frac{\Delta}{2} \sigma_z  &+& \sum_{k} \omega_k a^{\dagger}_{k}a_{k}  - i \sum_{k} g_{k} \left( a^{\dagger}_{k} \sigma_- - a_{k} \sigma_+ \right) \nonumber \\
&+& \frac{\Omega_x(t)}{2}(\sigma_+ + \sigma_-). 
\label{eq:hamiltonian_1}
\end{eqnarray}

where $\Delta=\omega_1-\omega_0$ is the detuning of the emitter from the target frequency, caused by the random fluctuation in the local strain or charge environment; this detuning is assumed to be static on the spontaneous emission timescale. The operators $\sigma_z=|e\rangle\langle e|-|g\rangle\langle g|$, $\sigma_+ =|e\rangle\langle g|$, and $\sigma_- = |g\rangle\langle e| = (\sigma_+)^\dagger$ are respectively, the $z$-axis Pauli matrix, the raising and the lowering operator for the two-level system. $a_k$ is the annihilation operator of the $k$-th photon mode, $g_k$ is its coupling strength to the emitter, and $\omega_k$ is its detuning from $\omega_0$.  We consider pulses such that $\Omega_x(t) = \Omega_x$ during the pulses and zero otherwise. We will further assume $\Omega_x$ to be much larger than all other relevant energy scales so that the pulses are essentially instantaneous (i.e. $\Omega_x \gg \Delta, \Gamma, g_k$ and $t_p \to 0$). At the initial time, $ t = 0$, we will assume that both emitters have their excited states occupied and ground states unoccupied and that all radiation modes are empty. Furthermore, we assume that both emitters are driven by identical pulse sequences.

In the setup described schematically by Fig. \ref{fig:HOM}, with the addition of an identical pulse sequence applied to both emitters, we want to evaluate the second order coherence equivalent to the intensity correlation at the detectors that is defined by:

\begin{equation}
\label{eq:G2_34_def}
 G^{(2)}_{34}(t,\theta) = \langle a_3^{\dagger}(t) a_4^{\dagger}(t+\theta)a_4(t+\theta) a_3(t) \rangle.
\end{equation}

From this, we will extract the integrated intensity correlation corresponding to the experimentally measured cross-correlation in the Hanburry Brown and Twiss setup\cite{HanburryBrownTwiss_1956, KirazAtature_PRA_2004}:
\begin{equation}
\label{eq:g2_34_integrated}
 g^{(2)}_{34}(\theta) = \lim_{T \to \infty} \int_0^T G_{34}^{(2)}(t, \theta) \; dt
\end{equation}

For a 50:50 beam splitter, the operators $a_3$ and $a_4$ at the detectors can be expressed in terms of the operators $a_1$ and $a_2$ at the emitters as:
\begin{eqnarray}
a_3(t) &=& \frac{1}{\sqrt{2}}\left( a_1(t) +i a_2(t) \right) \\
a_4(t) &=& \frac{1}{\sqrt{2}}\left( i a_1(t) + a_2(t) \right)
\end{eqnarray}
and similarly for their conjugate expressions. {\textit{k}}-integrated operators are used because they correspond to the electric field operators. Plugging this into the equations for the coherence, we get after dropping negligible two-photon terms:
\begin{eqnarray}
\label{eq:G2_34_emitters}
 G^{(2)}_{34}(t,\theta) &=& \frac{1}{4} \left \{ \langle a_1^{\dagger}(t) a_1(t)a_2(t+\theta) a_2^{\dagger}(t+\theta) \rangle \right. \nonumber \\
                  &+& \langle a_2^{\dagger}(t) a_2(t) a_1^{\dagger}(t+\theta) a_1(t+\theta) \rangle \nonumber \\
                  &-& \langle a_2^{\dagger}(t) a_2(t + \theta) a_1^{\dagger}(t+\theta) a_1(t) \rangle \nonumber \\
                  &-& \left. \langle a_1^{\dagger}(t) a_1(t+\theta) a_2^{\dagger}(t+\theta) a_2(t) \rangle \right\}
\end{eqnarray}
If we define $g_i(t,\theta) = \langle a^{\dagger}_i(t) a_i(t+ \theta) \rangle$ with $i = 1, 2$,  we can rewrite $G_{34}(t, \theta)$ as:
\begin{eqnarray}
\label{eq:G2_34_g}
G_{34}^{(2)}(t, \theta) &=& \frac{1}{4} \{ g_{1}(t,0) g_{2}(t+\theta,0) + g_{2}(t,0) g_{1}(t+\theta,0) \nonumber \\
&-& g^*_{1}(t, \theta) g_{2}(t,\theta) - g^*_{2}(t, \theta) g_{1}(t, \theta) \}. 
\end{eqnarray}

\begin{figure}[ht]
\includegraphics[height=6.50cm, width=8.0cm]{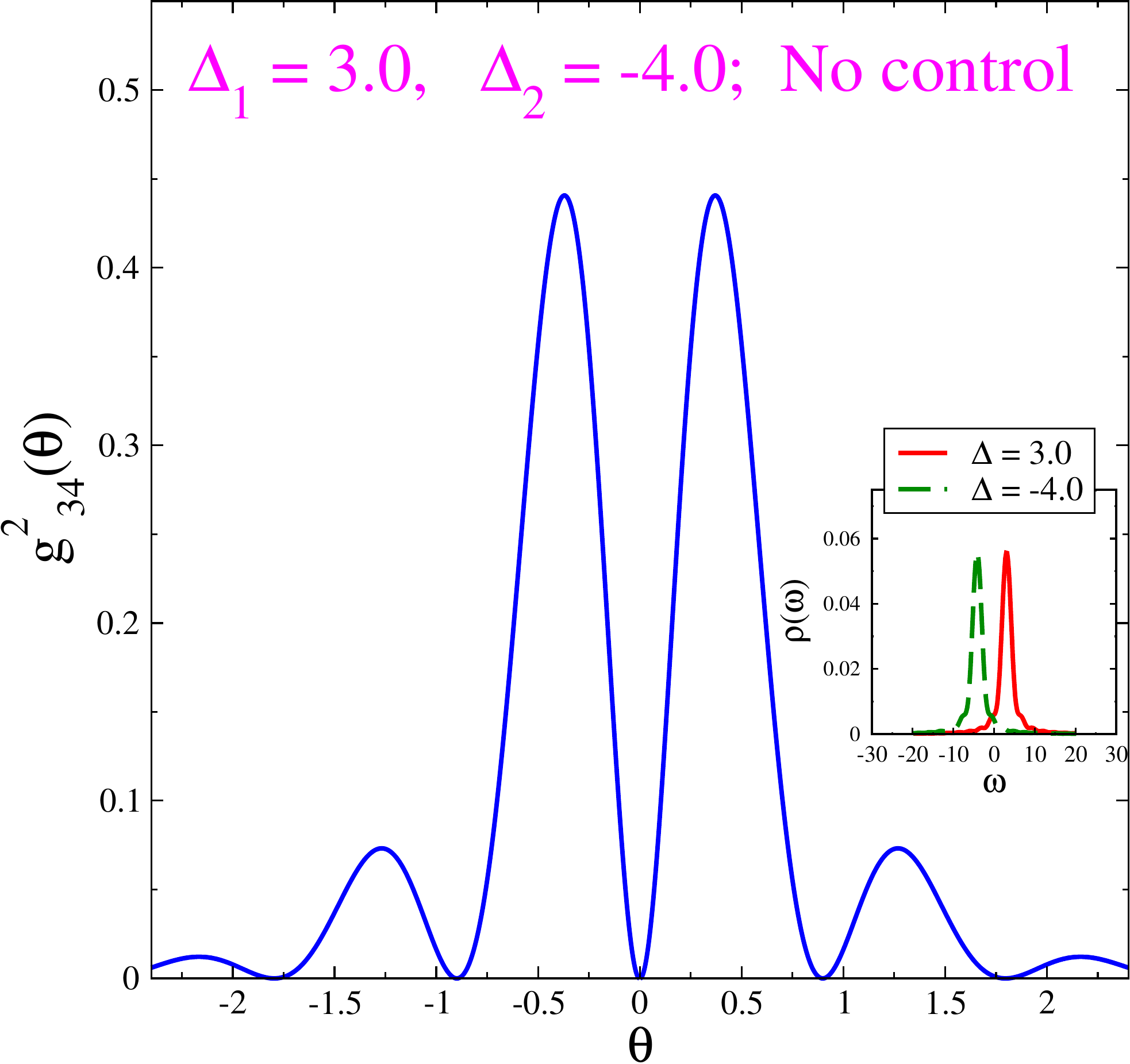}
\caption{Intensity correlation at the detectors in the absence of a control protocol with two independent emitters with detuning $\Delta_1 = 3.0$ and $\Delta_2 = -4.0$ respectively. The insert shows the emission spectra of individual emitters with detunings $\Delta_1 = 3.0$ (solid red line) and $\Delta_2 = -4.0$ (dashed green line). In the absence of any control protocol, they both have Lorentzian lineshapes centered around their respective detunings\cite{RF_Heitler_Book1960} and have limited spectral overlap.The intensity correlation vanishes at $\theta = 0$ and periodically at $\theta = 2n\pi/\Delta_{21}$, $n$ integer.}
\label{fig:g2_spectra_Deltas_3p0_m4p0_NoPulse}
\end{figure}

Calculating the second-order coherence then reduces to evaluating the first order coherences $g_i(t, 0)$ and $g_i(t, \theta)$.  For this, we will use the master equations characterizing the time-evolution of the density matrix operator for individual emitters: $\rho(t) = \rho_{ee}(t) |e\rangle \langle e| + \rho_{eg}(t) |e\rangle \langle g| + \rho_{ge}(t) |g \rangle \langle e|  + \rho_{gg}(t) |g\rangle \langle g| $ with $\rho_{ge}^* = \rho_{eg}$ and $\rho_{ee} + \rho_{gg} = 1$.

\section{Pulse-Driven Emitters and Solution}
The master equations or optical Bloch equations governing the time-evolution of the above-defined density matrix operator is given in the rotating wave approximation by~\cite{Cohen_Tannoudji_Book1992}:
\begin{equation}
\label{eq:MasterEquation_1} 
\begin{split}
\dot{\rho}_{ee} &= i \frac{\Omega_x(t)}{2}(\rho_{eg} - \rho_{ge}) - \Gamma \rho_{ee} \; ,   \\
\dot{\rho}_{gg} &= -i \frac{\Omega_x(t)}{2}(\rho_{eg} - \rho_{ge}) + \Gamma \rho_{ee} \; ,  \\
\dot{\rho}_{ge} &= -i \frac{\Omega_x(t)}{2}\left(\rho_{ee} - \rho_{gg}\right) + (i\Delta -\frac{\Gamma}{2}) \rho_{ge}  \; , \\
\dot{\rho}_{eg} &=  i \frac{\Omega_x(t)}{2}\left(\rho_{ee} - \rho_{gg}\right) + (-i\Delta -\frac{\Gamma}{2}) \rho_{eg} \; .     
\end{split}
\end{equation}

where $\Gamma = 2\pi\int g_k^2 \; \delta(\omega_k - \Delta) \; dk$ is the spontaneous emission rate in the absence of the control field. It is used to set the units of time and energy. Namely, we set $\Gamma = 2$ and we measure energy in units of $\Gamma$ and time in units of $1/\Gamma$. 
Each $\pi_x$ pulse inverts the populations of the excited and ground state and swaps the values of $\rho_{eg}$ and $\rho_{ge}$. 

In the far field region, the field operators, $a$ and $a^{\dagger}$, are related to the emitter operators, $\sigma_+$ and $\sigma_-$, by simple proportionality constants independent of time. We can accordingly redefine the coherences so that\cite{Loudon_book1983}: 
\begin{equation}
 g_i(t, \theta) = \langle \sigma_{+ i}(t) \sigma_{- i}(t + \theta) \rangle
\end{equation}
The two-time correlation function $\langle \sigma_+(t) \sigma_-(t+\theta) \rangle$ can be  expressed as a single-time expectation value according to the quantum regression theorem or following~\cite{RF_Mollow_PhysRev1969, Scully_Zubairy_book1997, Loudon_book1983, Scully_Zubairy_book1997, Fotso_JPhysB2019}:
\begin{eqnarray}
 &&\hspace{2.250cm}\langle \sigma_+(t) \sigma_-(t+\theta) \rangle  \nonumber \\
 &=&\mathrm{Tr} \left[ \rho(0)U^{\dagger}(0,t)\sigma_+ U(0,t)U^{\dagger}(0,t+\theta)\sigma_-U(0,t+\theta)\right] \nonumber \\
 & &    \\ 
 &=& \mathrm{Tr} \left[ U(t,t+\theta) \rho(t) \sigma_+ U^{\dagger}(t,t+\theta) \sigma_- \right] \\
 &=& \mathrm{Tr} \left[ \rho^{\prime}(t + \theta) \sigma_- \right] 
 \label{eq:expVal_sigmaP_sigmaM}
\end{eqnarray}

where $\rho^{\prime}(t) =  \rho(t) \sigma_+ $, and where $\sigma_+$ and $\sigma_-$ are the time-independent operators in the Schr\"odinger picture. $U(t,t')$ is the time-evolution operator from time $t$ to $t'$ for the system described by Eqs.(\ref{eq:MasterEquation_1}). 

With the assumption that each emitter is initially prepared in its excited state and all bosonic modes are initially empty, the expression in Eq.~\ref{eq:expVal_sigmaP_sigmaM} can be evaluated by integrating numerically or analytically the master equation between consecutive pulses.

\begin{figure}[t] 
\includegraphics[height=6.50cm, width=8.0cm]{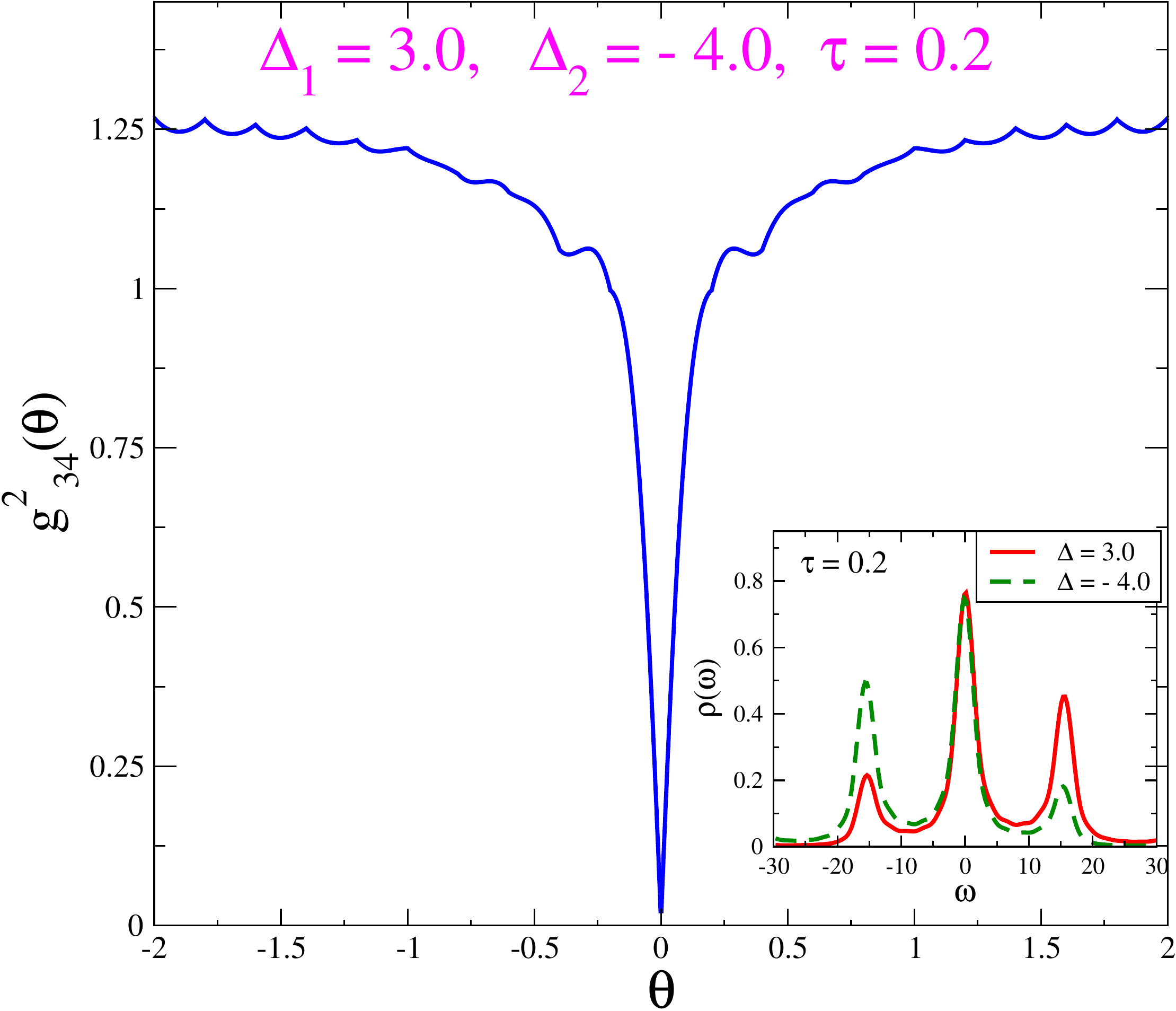}
\caption{Intensity correlation at the detectors when the two emitters, with detunings $\Delta_1 = 3.0$ and $\Delta_2 = -4.0$ respectively, are driven by a periodic sequence of $\pi$-pulses with period $\tau = 0.2$. The insert shows the emission spectra of individual emitters with detunings $\Delta_1 = 3.0$ (solid red line) and $\Delta_2 = -4.0$ (dashed green line) when they are driven by the same periodic pulse sequence\cite{FotsoEtal_PRL2016}. The intensity correlation is zero at $\theta = 0$ and finite elsewhere.}
\label{fig:g2_spectra_Deltas_3p0_m4p0_Tau0p2}
\end{figure}

\subsection{No control protocol}
In the absence of a control protocol, a straightforward integration of the equations yields:
\begin{equation}
 g_{1, \; 2}(t, \theta) = \mathrm{e}^{-\Gamma t} \mathrm{e}^{ (- \Gamma/2 + i\Delta_{1, \; 2} )\theta}.
\end{equation}
Plugging this into Eq.(~\ref{eq:G2_34_g}), we get: 
\begin{eqnarray}
 G^{(2)}_{34}(t, \theta) & = & \frac{1}{4} \left \{ \mathrm{e}^{-\Gamma t} \mathrm{e}^{ - \Gamma(t + \theta)} + \mathrm{e}^{-\Gamma t} \mathrm{e}^{ - \Gamma(t + \theta)} \right. \nonumber \\
                  & - & \mathrm{e}^{-\Gamma t} \mathrm{e}^{ (-\Gamma/2 - i \Delta_2)\theta }  \mathrm{e}^{- \Gamma t} \mathrm{e}^{(-\Gamma/2 + i \Delta_1)\theta} \nonumber \\
                  & - &  \left. \mathrm{e}^{-\Gamma t} \mathrm{e}^{(-\Gamma/2 - i \Delta_1)\theta } \mathrm{e}^{-\Gamma t} \mathrm{e}^{(-\Gamma/2 + i \Delta_2)\theta } \right\} \nonumber \\
                  & = & \frac{1}{2} \mathrm{e}^{-2\Gamma t - \Gamma \theta} \left[ 1 - \mathrm{cos} \Delta_{21} \theta \right] 
\end{eqnarray}
with $\Delta_{21} = \Delta_2 - \Delta_1$.
From this, we obtain:
\begin{eqnarray}
\label{eq:g2_34_integrated_noPulse}
 g^{(2)}_{34}(\theta) & = &  \int_0^T G_{34}^{(2)}(t, \theta) \; dt  \nonumber \\
 & = & \frac{1}{4\Gamma}\left(1 - \mathrm{e}^{-2\Gamma T} \right) \mathrm{e}^{- \Gamma \theta} \left(1 - \mathrm{cos}(\Delta_{21} \theta) \right).
\end{eqnarray}
For $T \to \infty$, this simplifies to:
\begin{equation}
\label{eq:g2_34_integrated_noPulse_longTime}
 g^{(2)}_{34}(\theta)  = \frac{1}{4\Gamma} \left(1 - \mathrm{cos}(\Delta_{21} \theta) \right) \mathrm{e}^{- \Gamma \theta}.
\end{equation}
This is a function that takes an identical minimum at zero time delay and repeatedly in a periodic way with a period defined by $\theta = 2n\pi/\Delta_{21}$, with $n$ integer..

\begin{figure}[htbp]
\includegraphics[height=2.50cm, width=7.50cm]{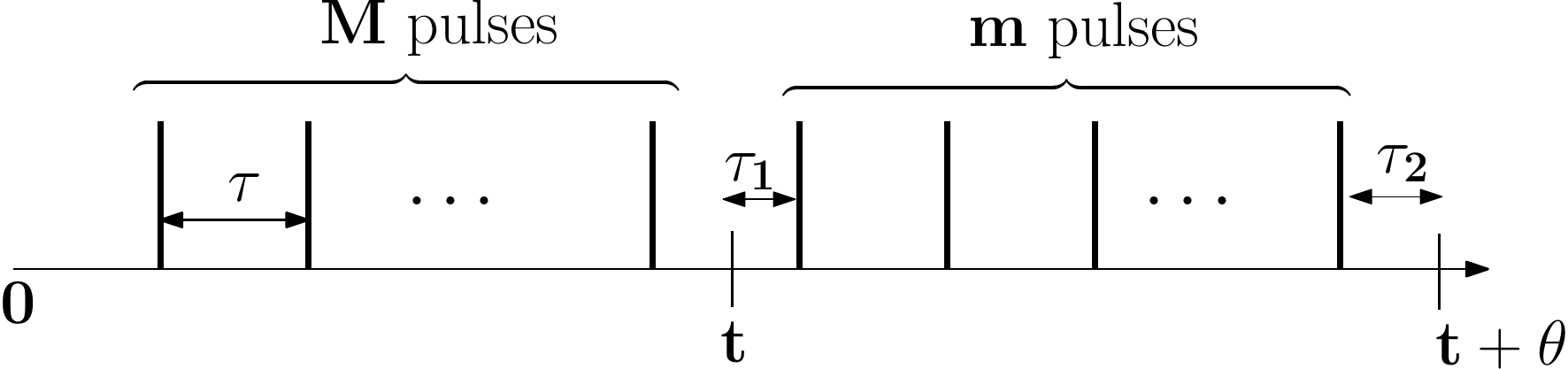}
\caption{Schematic representation of the detection times $t$ and $t + \theta$ on the time axis.}
\label{fig:t_t_plus_theta}
\end{figure}

\subsection{Periodic pulse sequence}
For a periodic pulse sequence, the evaluation of $g_i(t, \theta)$ is achieved by iteratively integrating Eqs.(~\ref{eq:MasterEquation_1}) between consecutive pulses, using the fact that the effect of each pulse is to swap the populations of the excited and ground states as well as the coherences. This integration follows steps similar to those highlighted in Refs.\cite{FotsoDobrovitski_Absorption, Fotso_JPhysB2019} and yields:
\begin{equation}
 g(t, \theta) = f(t, \theta) \rho_{gg}(t).
\end{equation}
With the times $t$ and $t+\theta$ such that $t = M\tau + (\tau - \tau_1)$ and $t + \theta = (M+m)\tau + \tau_2$ as illustrated schematically in Fig.\ref{fig:t_t_plus_theta}, $\rho_{gg}(t)$ is given by:
\begin{equation}
\label{eq:rho_gg}
 \rho_{gg}(t) = 1 - \frac{ 1 - (-1)^{M+1} \mathrm{e}^{ -(M+1)\Gamma\tau}}{1 + \mathrm{e}^{-\Gamma\tau}} \mathrm{e}^{-\Gamma (\tau -\tau_1)}.
\end{equation}

\begin{figure}[htbp]
\includegraphics[height=6.50cm, width=8.0cm]{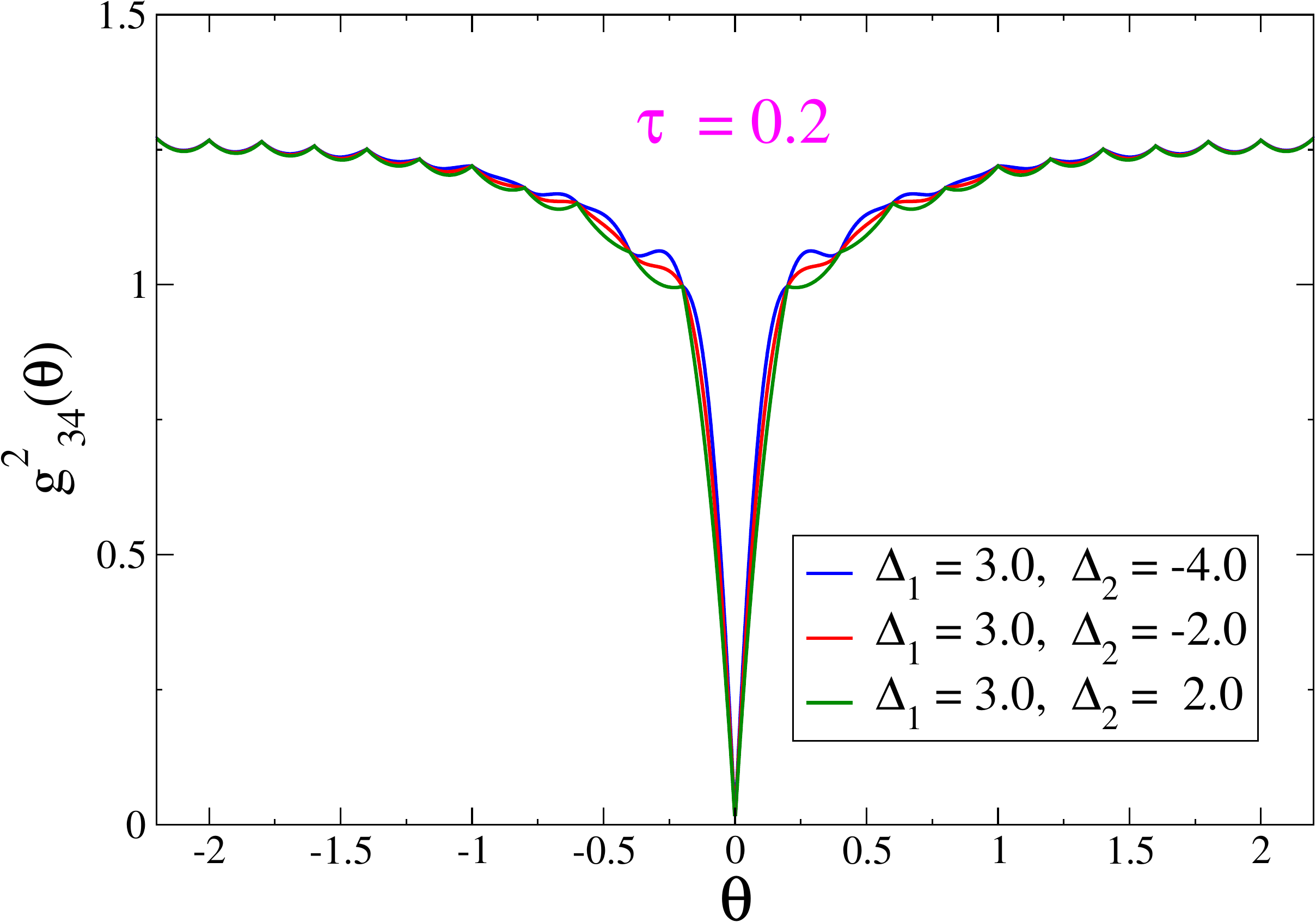}
\caption{Intensity correlation at the detectors when the two emitters, with detunings $\Delta_1 = 3.0$ and $\Delta_2 = -4.0$ (blue line),  $\Delta_1 = 3.0$ and $\Delta_2 = -2.0$ (red line), $\Delta_1 = 3.0$ and $\Delta_2 = 2.0$ (green line), are driven by a periodic sequence of $\pi$-pulses with period $\tau = 0.2$.}
\label{fig:g2_spectra_3Deltas_tau0p2}
\end{figure}

\begin{figure}[htbp]
\includegraphics[height=6.50cm, width=8.0cm]{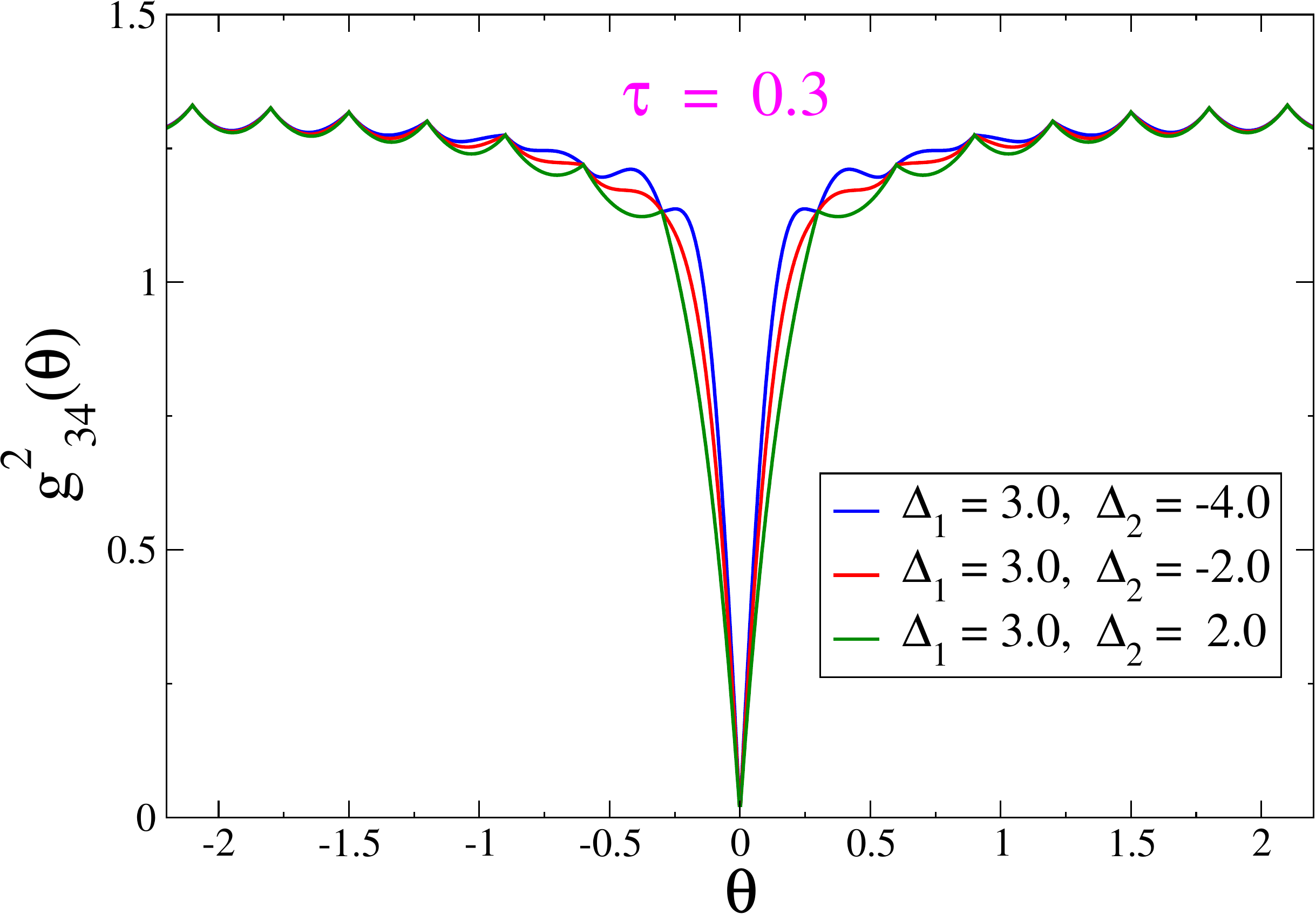}
\caption{Intensity correlation at the detectors when the two emitters, with detuning $\Delta_1 = 3.0$ and $\Delta_2 = -4.0$ (blue line),  $\Delta_1 = 3.0$ and $\Delta_2 = -2.0$ (red line), $\Delta_1 = 3.0$ and $\Delta_2 = 2.0$ (green line), are driven by a periodic sequence of $\pi$-pulses with period $\tau = 0.3$.}
\label{fig:g2_spectra_3Deltas_tau0p3}
\end{figure}

The function $f(t, \theta)$ is such that:
\begin{itemize}
 \item For $t$ and $t+\theta$ in the same pulse interval, we have:
 \begin{equation}
 \label{eq:f_theta_1}
  f(t, \theta) = \mathrm{e}^{\left( i \Delta - \frac{\Gamma}{2}\right)\theta};
 \end{equation}

 \item For $t$ and $t+\theta$ separated by an odd number of pulse intervals, we have:
 \begin{equation}
 \label{eq:f_theta_2}
  f(t, \theta) = 0;
 \end{equation} 
 
 \item For $t$ and $t+\theta$ separated by an even number of pulse intervals, we have:
 \begin{equation}
 \label{eq:f_theta_3}
  f(t, \theta) = \mathrm{e}^{ -\Gamma \theta/2 }  \mathrm{e}^{ i \Delta \left(\theta - m\tau\right)}.
 \end{equation} 
\end{itemize}

These expressions, combined with those of $g(t, 0) = \rho_{gg}(t)$ and $g(t+\theta, 0) = \rho_{gg}(t + \theta)$ can be brought into the expression of $G^{(2)}_{34}(t, \theta)$ and the integral for $g^{(2)}_{34}(\theta)$ completed numerically.

\section{Results}

Fig.\ref{fig:g2_spectra_Deltas_3p0_m4p0_NoPulse} and Fig.\ref{fig:g2_spectra_Deltas_3p0_m4p0_Tau0p2} are the main results of this paper. They present, for two emitters with detunings $\Delta_1 = 3.0$ and $\Delta_2 = -4.0$, the intensity correlation at the detectors in the absence of a control protocol and in the presence of a periodic pulse sequence of period $\tau = 0.2$ respectively. The inserts show in both situations the emission spectra of the individual emitters. In the absence of a control protocol, the emission spectra have Lorentzian lineshapes centered around the emitters detunings (in the frame rotating at $\omega_0$). Thus, in this situation, the spectral overlap is limited for $ |\Delta_{21}| = |\Delta_1 - \Delta_2| > \Gamma$. In this case, the intensity correlation vanishes periodically with a period defined by the value $\Delta_{21}$ of the difference between the frequencies of the two emitters\cite{Legero_Kuhn_PRL2004}, namely, it vanishes at $\theta = 2n\pi/\Delta_{21}$, $n$ integer.

In the second situation, i.e when the two emitters are driven by the same periodic pulse sequence of period $\tau=0.2$, as shown in the insert, spectral overlap is enhanced with both emission spectra having overlapping central peaks despite $|\Delta_{21}| > \Gamma$. The intensity correlation vanishes at $\theta = 0$ and remains finite for $\theta \ne 0$ indicating enhanced photon indistinguishability.

In Fig.\ref{fig:g2_spectra_3Deltas_tau0p2} and Fig.\ref{fig:g2_spectra_3Deltas_tau0p3}, we show that the intensity correlations for $\tau = 0.2$ and for $\tau=0.3$ respectively, for $\Delta_1 = 3.0$ and $\Delta_2 = -4.0$ (blue line),  $\Delta_1 = 3.0$ and $\Delta_2 = -2.0$ (red line), $\Delta_1 = 3.0$ and $\Delta_2 = 2.0$ (green line). It clearly exhibits little dependence on the individual emitters detunings. Furthermore, these figures show the dependence of the intensity correlations on the pulse period $\tau$. The function vanishes at $\theta = 0$ and remains finite elsewhere. Note that the small oscillations away from the minimum correspond to beating at the pulse sequence period. Note that the intensity correlation is nearly identical for $\Delta_{21}$ spanning a range of width $\sim 10\Gamma$. These figures demonstrate an enhanced photon indistinguishability as long as $\Delta_{1,2} \; \times \tau \le 1$. 

The results presented here are obtained using the transient expression of $\rho_{gg}$ (Eq. \ref{eq:rho_gg}) and $f(t, \theta)$ (Eqs. \ref{eq:f_theta_1}, \ref{eq:f_theta_2}, \ref{eq:f_theta_3}) for a total time equal to $4.8$ but one could also use the long time stationary regime where $\rho_{gg} \approx 1/2 $ under a periodic pulse sequence. In this case, we will get:
\begin{eqnarray}
 G^{(2)}_{34}(t, \theta) & = & \frac{1}{4} \left[ \frac{1}{2} \times \frac{1}{2} + \frac{1}{2} \times \frac{1}{2} \right. \nonumber \\
                   & - & \left. \frac{1}{4} f_2^*(t, \theta) f_1(t, \theta) - \frac{1}{4} f_1(t, \theta) f_2^*(t, \theta) \right] \\
     & = & \frac{1}{8} \left[ 1 - \mathrm{Real}\left\{ f_1(t, \theta) f_2^*(t, \theta) \right\} \right]  
\end{eqnarray}
The results are overall similar with the only difference that the beating at long times is strongly suppressed in the steady state regime.\\

\section{Conclusion}
Photon indistinguishability is essential for a variety of photon-mediated operations in quantum information processing. For quantum emitters in the solid state or in other dynamic environments, this can be compromised by spectral diffusion due to fluctuations in the surrounding bath of the emitter resulting in low efficiency for the aforementioned operations. We have examined the two-photon interference in the context of a HOM-type experiment when the two involved quantum emitters are subject to spectral diffusion that reduces the spectral overlap. We have evaluated the intensity correlation at the detectors when both emitters are driven by a periodic sequence of $\pi$ pulses of period $\tau$. In the absence of the control protocol, the intensity correlation exhibits beating with vanishing values periodically as a function of the delay time $\theta$ between the two the detectors. Under the control sequence, the intensity correlation vanishes at zero time delay but remains finite otherwise. These results indicate that the control protocols can indeed enhance photon indistinguishability and thus improve the efficiency of fundamental operations that are central to the development of scalable quantum information processing and quantum networks for qubits susceptible to spectral diffusion.



\end{document}